\documentclass{ws-p10x7}

\begin{document}

\title{Recent KTEV Results}

\author{R.Kessler}

\address{Enrico Fermi Institute, University of Chicago \\
         5640 South Ellis Avenue, Chicago Illinois 60637, USA  \\
         E-mail: kessler@hep.uchicago.edu
            }

\def\epe{\Re(\epsilon^{\prime}/\epsilon) }
\def\imepe{\Im(\epsilon^{\prime}/\epsilon) }

\twocolumn[\maketitle\abstract{
  Preliminary KTEV results are presented based on the 1997 data set,
  and include an improved measurement of $\epe$,
  CPT tests, and precise measurements of $\tau_S$ and $\Delta m$.
}]


\def\ll{\ell^+\ell^-}
\def\llbar{\ell\bar{\ell}}
\def\nn{\nu\bar{\nu}}
\def\pp{\pi^+\pi^-}
\def\mm{\mu^+\mu^-}
\def\ee{e^+e^-}
\def\gg{\gamma\gamma}
\def\firstobs{$(1^{st}~obs)$ }
\newcommand\kgain[1]{$(\times #1)$}

\def\pipi{\pi^+\pi^-}
\def\pzpz{\pi^0\pi^0}
\def\kthree{K_L\to 3\pi^0}
\def\kneut{K_{L,S}\to \pi^0\pi^0}
\def\kchrg{K_{L,S}\to \pi^+\pi^-}
\def\kpmz{K_{L}\to \pi^+\pi^-\pi^0}
\def\eu{\times 10^{-4}}
\def\reg{regenerator}
\def\bkg{background}
\def\dke3{\delta_{Ke3}}
\def\kbar{\overline{K^0}}
\def\ke3{K_L\to\pi e \nu}
\def\pt2{p_t^2}

\def\delm{\Delta m}
\def\delgam{\Delta\Gamma}
\def\delphi{\Delta\Phi}
\def\phipm{\Phi_{+-}}
\def\phizz{\Phi_{00}}

\def\CPTV{\begin{picture}
             (25,9) \put(0,0) {$CPT$} \put(0,0) {\line(3,1){22}}
          \end{picture}}

\def\CPV{\begin{picture}
             (25,9) \put(0,0) {$CP$} \put(0,0) {\line(3,1){18}}
          \end{picture}}


            \section{Introduction}


The KTeV experiment at Fermilab was designed primarily to 
measure the direct CP violating parameter $\epe$ with much better
precision than previous measurements at 
Fermilab\cite{e731} and CERN\cite{na31} in the early 1990's.
The measured quantity is the double ratio of decay rates,
\begin{equation}
 {\cal R} \equiv  
  {
    { \Gamma(K_L\to \pi^+\pi^-) / \Gamma(K_S\to \pi^+\pi^-) }
             \over
    { \Gamma(K_L\to \pi^0\pi^0) / \Gamma(K_S\to \pi^0\pi^0) }
  }
\end{equation}
where ${\cal R} \simeq 1 + 6\epe$.
The CP-violating $K_L\to 2\pi$ decays can be explained 
by the parameter $\epsilon \sim 2\times 10^{-3}$,
which describes the tiny CP-violating asymmetry in the
$K^0~\leftrightarrow~\overline{K^0}$ mixing.
If the double ratio ${\cal R}$ differs from unity, i.e., $\epe \ne 0$, 
this would be a clear indication of direct CP-violation
in the decay amplitude.  In short, ${\cal R} \ne 1$
implies that
\begin{eqnarray} 
      A(K^0\to \pipi) & \ne & \bar{A}(\overline{K^0}\to \pipi) \nonumber \\
      A(K^0\to \pzpz) & \ne & \bar{A}(\overline{K^0}\to \pzpz) \nonumber
\end{eqnarray}
which would demonstrate that a particle and its 
anti-particle partner can decay differently.
Note that CPT requires 
$\sum \vert A_i\vert^2 = \sum \vert \bar{A}_i\vert^2$.

The experimental challenge is to measure the double ratio ${\cal R}$
based on millions of decays, and then to control the systematic
uncertainties to much better than a percent.
The key is to collect $K_L$ and $K_S$ decays at the same time
so that common charged or neutral mode systematic uncertainties
will cancel in the $K_L/K_S$ single ratios. 


     \section{Measurement Technique}

\begin{figure*}[ht]
\epsfxsize400pt
\figurebox{160pt}{160pt}{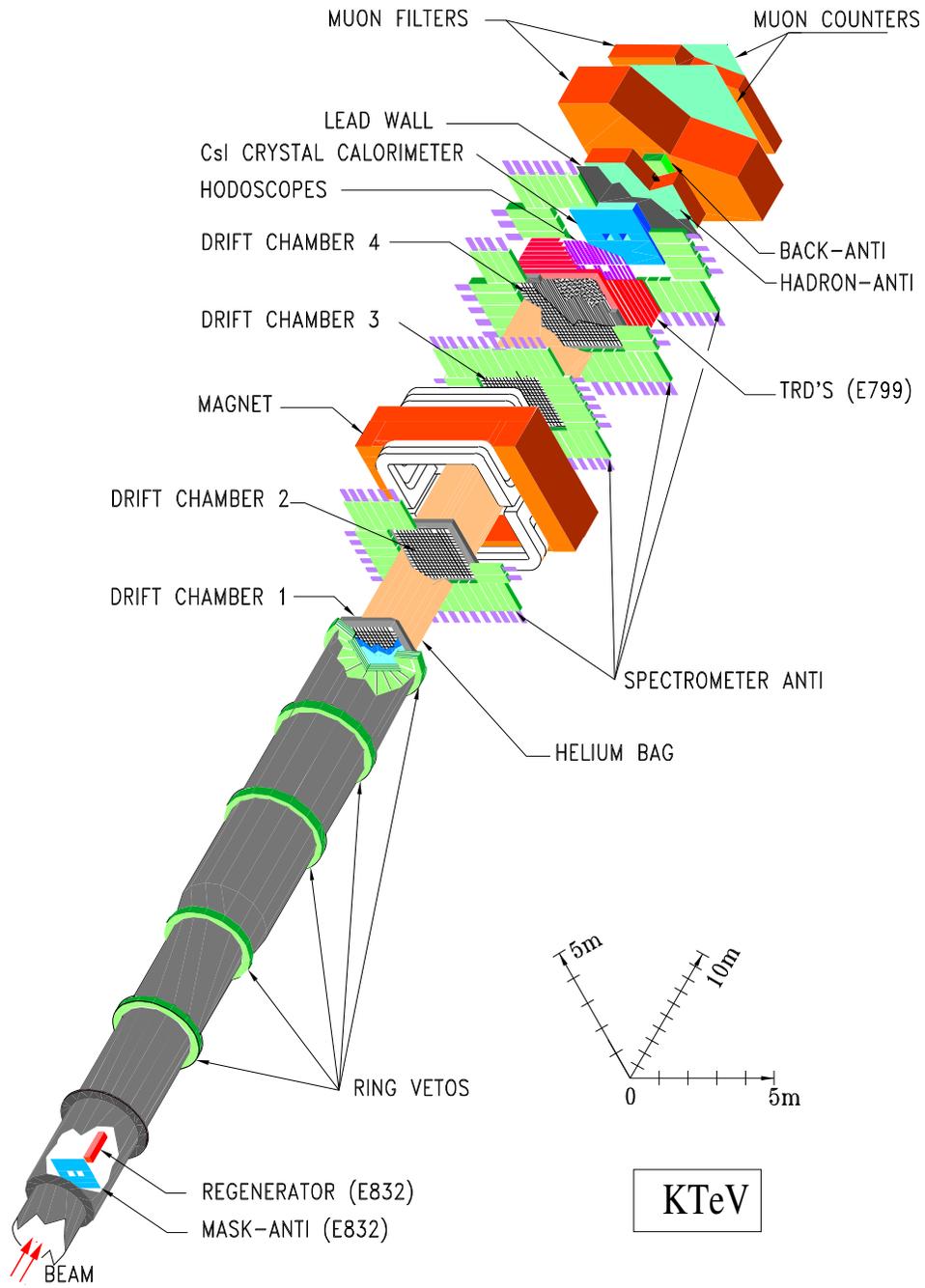}
\caption{KTEV Detector}
\label{fig:det3d}
\end{figure*}
\begin{figure*}[ht]
\epsfxsize400pt
\figurebox{160pt}{160pt}{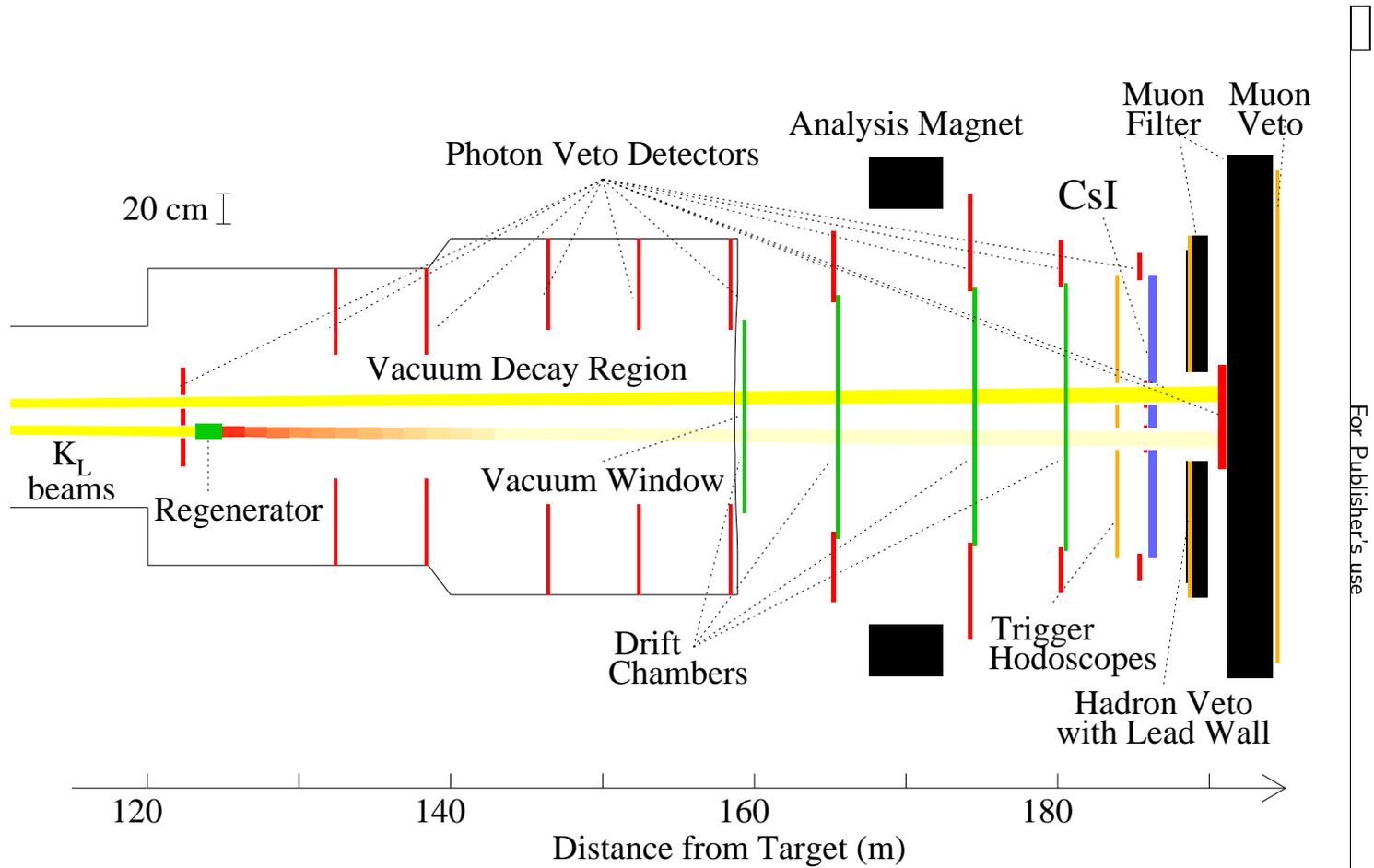}
\caption{Plan view of KTEV Detector}
\label{fig:det2d}
\end{figure*}

The KTEV detector is shown in Figures~\ref{fig:det3d}-\ref{fig:det2d}.
Kaons were produced by 800 GeV protons hitting a 50~cm long
BeO target at 4.8~mrad target angle.
Two nearly parallel $K_L$ beams entered the decay region
roughly 120 meters from the primary production target.
One of the beams hit a 1.8 meter long {\reg} made of plastic
scintillator; the kaon state exiting the {\reg} was
$K_L + \rho K_S$, where $\rho \sim 0.03$ was sufficient 
so that ``{\reg}'' $2\pi$ decays were dominated by $K_S\to 2\pi$.
The vacuum decay region extends up to 159 meters.
The $\kchrg$ decays were detected by a spectrometer consisting of
a magnet (411 MeV/c kick in horizontal plane) with two drift chambers
on each side. The neutral decays were detected by a pure 3100 channel
CsI calorimeter located  186 meters from the primary target.
The neutral and charged detector performances are summarized below:

\begin{itemize}
  \item NEUTRAL (CsI):
     \begin{itemize}
        \item energy resolution:  0.7\% at 15~GeV  \\
              (1.3\% at 3~GeV)
        \item energy non-linearity (3-75 GeV): 0.4\%
        \item position resolution: $\sim 1$ mm
     \end{itemize}
  \item CHARGED:
      \begin{itemize}
          \item drift chamber resolution: $100~\mu m$
          \item momentum resolution ($p$ in GeV):
              $$ {\sigma_p \over p} = 
                   \left[ 1.7 \oplus {p\over 14} \right] \times 10^{-3} $$
      \end{itemize}
\end{itemize}

 \clearpage

\begin{figure}[ht]
\epsfxsize200pt
\figurebox{160pt}{160pt}{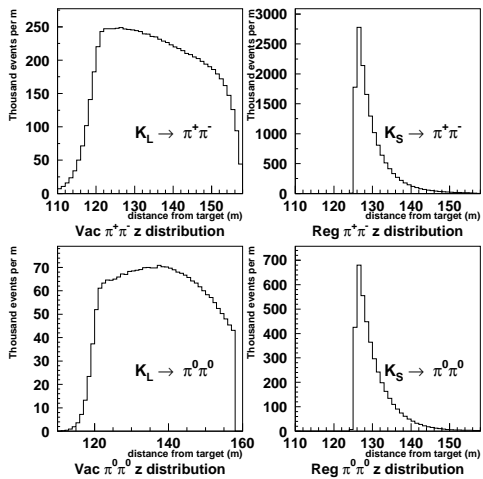}
\caption{Decay vertex distributions for $K_L$ and $K_S$ decays.}
\label{fig:vtxz}
\end{figure}

\begin{figure}[ht]
\epsfxsize200pt
\figurebox{160pt}{160pt}{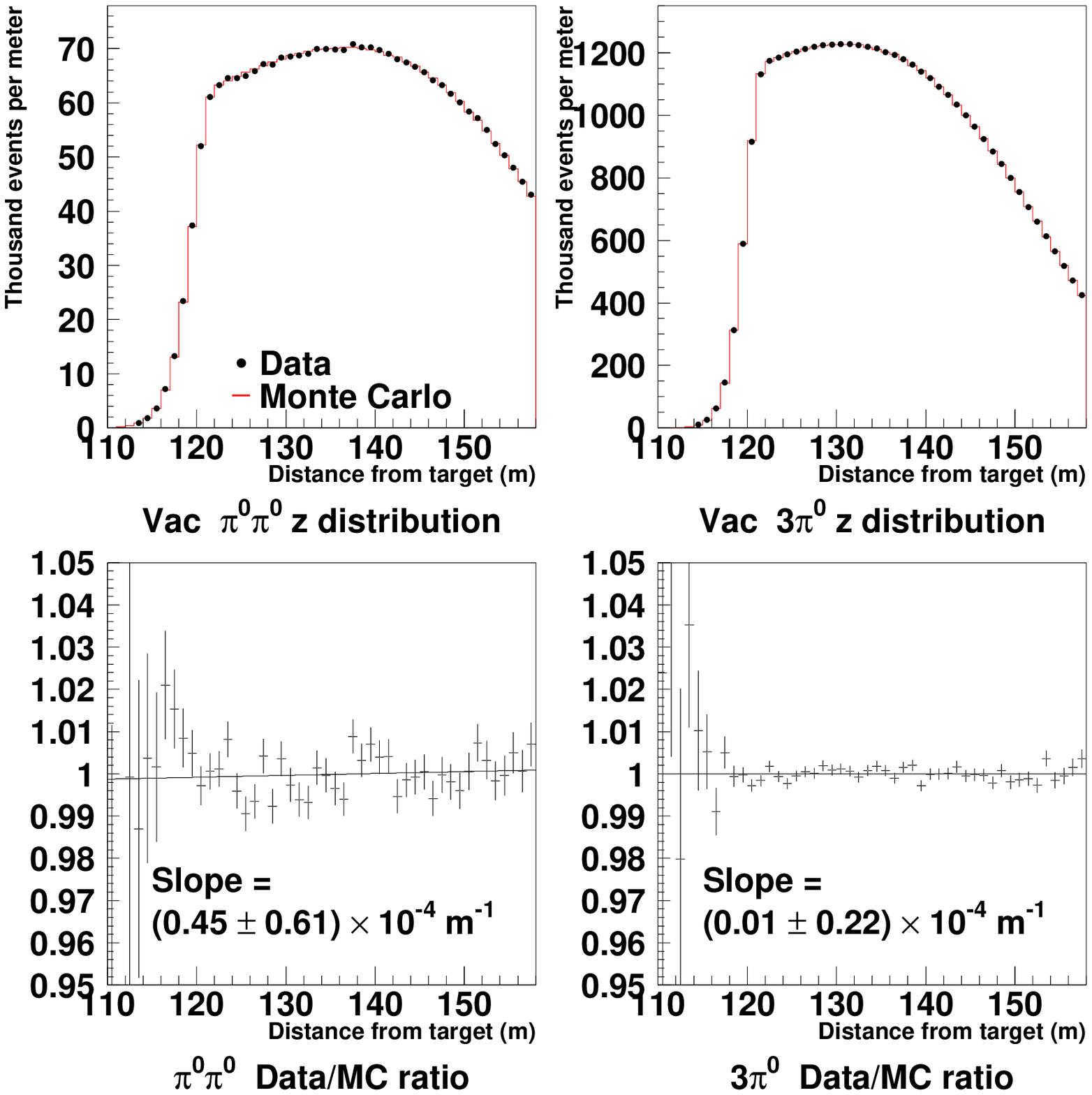}
\caption{Data/MC comparison of decay vertex for $K_L\to\pzpz$ (left)
         and for $\kthree$ (right). 
         The lower plots show the data/MC ratio vs. decay vertex. }
\label{fig:zneut97}
\end{figure}

\begin{figure}[ht]
\epsfxsize200pt
\figurebox{160pt}{160pt}{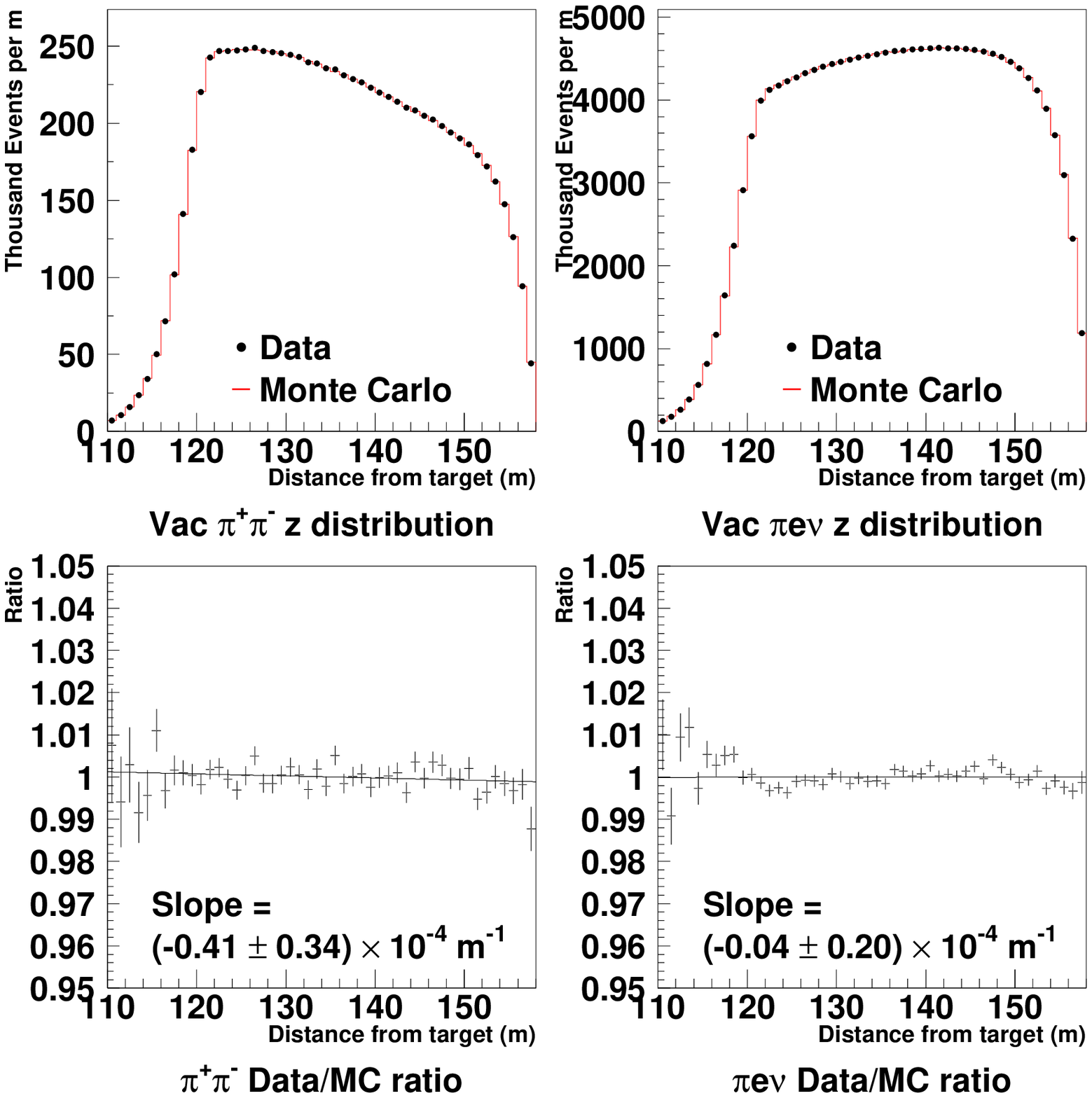}
\caption{Data/MC comparison of decay vertex for $K_L\to\pipi$ (left)
         and for $\ke3$ (right).
         The lower plots show the data/MC ratio vs. decay vertex. }
\label{fig:zchrg97}
\end{figure}

The $\pipi$ and $\pzpz$ mass resolutions are both $1.5~MeV/c^2$.
The decay distributions for $K_L$ and $K_S$ decays is shown in
Figure~\ref{fig:vtxz}. The difference in these two distributions
results in different acceptances that are corrected using a Monte Carlo
[MC] simulation. The quality of the simulation is shown by comparing
the data and MC decay vertex distributions, and is shown in 
Figures~\ref{fig:zneut97}-\ref{fig:zchrg97} for neutral and charged
decays in the $K_L$ beam.  There are 2.5 million $K_L\to\pzpz$ decays,
which agree very well with the 20 million MC events.
To further check our MC, 39 million $\kthree$ decays are compared
with an equal size MC sample, and the agreement is excellent.
In charged mode, 10 million $K_L\to\pipi$ decays agree very well
with the MC sample of 60 million. As a charged mode cross-check,
a sample of 170 million $\ke3$ decays are compared with an MC sample 
of 40 million; there is no $z$-slope in data/MC vs. decay vertex,
but there are non-statistical fluctuations suggesting subtle 
problems that might be related to the reconstruction with a 
missing neutrino.


  \section{ Regenerator Beam Results }

\begin{figure}[hb]
\epsfxsize200pt
\figurebox{160pt}{160pt}{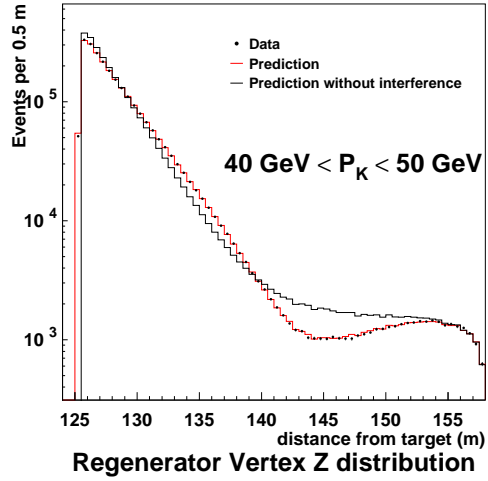}
\caption{Decay distribution in {\reg} beam.}
\label{fig:zreg}
\end{figure}

The $\kchrg$ decay distribution in the {\reg} beam 
is shown in Figure~\ref{fig:zreg}. The kaon momentum range 
40-50 GeV illustrates the high precision
with which we can measure parameters associated with the
interference.  These data (40-160 GeV) are used to make 
precision measurements
of $\tau_S$, $\delm$, $\delphi$ and $\phipm$.
The distribution of decays in the {\reg} beam is given by
\begin{eqnarray}
  {dN\over dt} 
     & = & \vert\rho\vert^2 e^{-\Gamma_S t} 
       +   \vert\eta\vert^2 e^{-\Gamma_L t} \label{eq:zfitfun} \\
     & + & 2\vert\rho\vert  \vert\eta\vert e^{(\Gamma_S+\Gamma_L)t/2}
               \cos(\delm t + \phi_{\rho} - \Phi_{\eta})   \nonumber
\end{eqnarray}
where $\Phi_{\eta} = \phipm,\phizz$ for charged,neutral decays.
For all fits described in the subsections below, 
the regeneration phase $\phi_{\rho}$ is determined 
from analyticity\cite{roy}.

     \subsection{ $\tau_S$ and $\delm$ }

The acceptance-corrected data was fit to the function in 
Eq.~\ref{eq:zfitfun}; $\tau_S$ and $\delm$ are floated and
CPT is assumed so that the phase of $\eta$ is given by
the superweak phase, $\Phi_{\eta} = \Phi_{SW} = \tan^{-1}(2\delm/\delgam)$.
The combined charged+neutral results are 
\begin{eqnarray}
   \tau_S & = & [89.67 \pm 0.04_{stat} \pm 0.04_{syst}]   \\
          &   &     psec \nonumber \\
   \delm  & = & [52.62 \pm 0.08_{stat} \pm 0.13_{syst} ]      \\
          &   &  \times 10^8~ \hbar s^{-1}  \nonumber
\end{eqnarray}
and are shown in 
Figures~\ref{fig:taus}-\ref{fig:delm}.
Our new $\tau_S$ value is consistent with previous measurements,
and it is $2.5\sigma$ above the PDG2000 value.
Similarly our new $\delm$ value is consistent with previous measurements,
and it is $2.1\sigma$ below the PDG2000 value.

  \subsection{CPT Tests: $\delphi$ and $\phipm$ }

\begin{figure}
\epsfxsize220pt
\figurebox{160pt}{160pt}{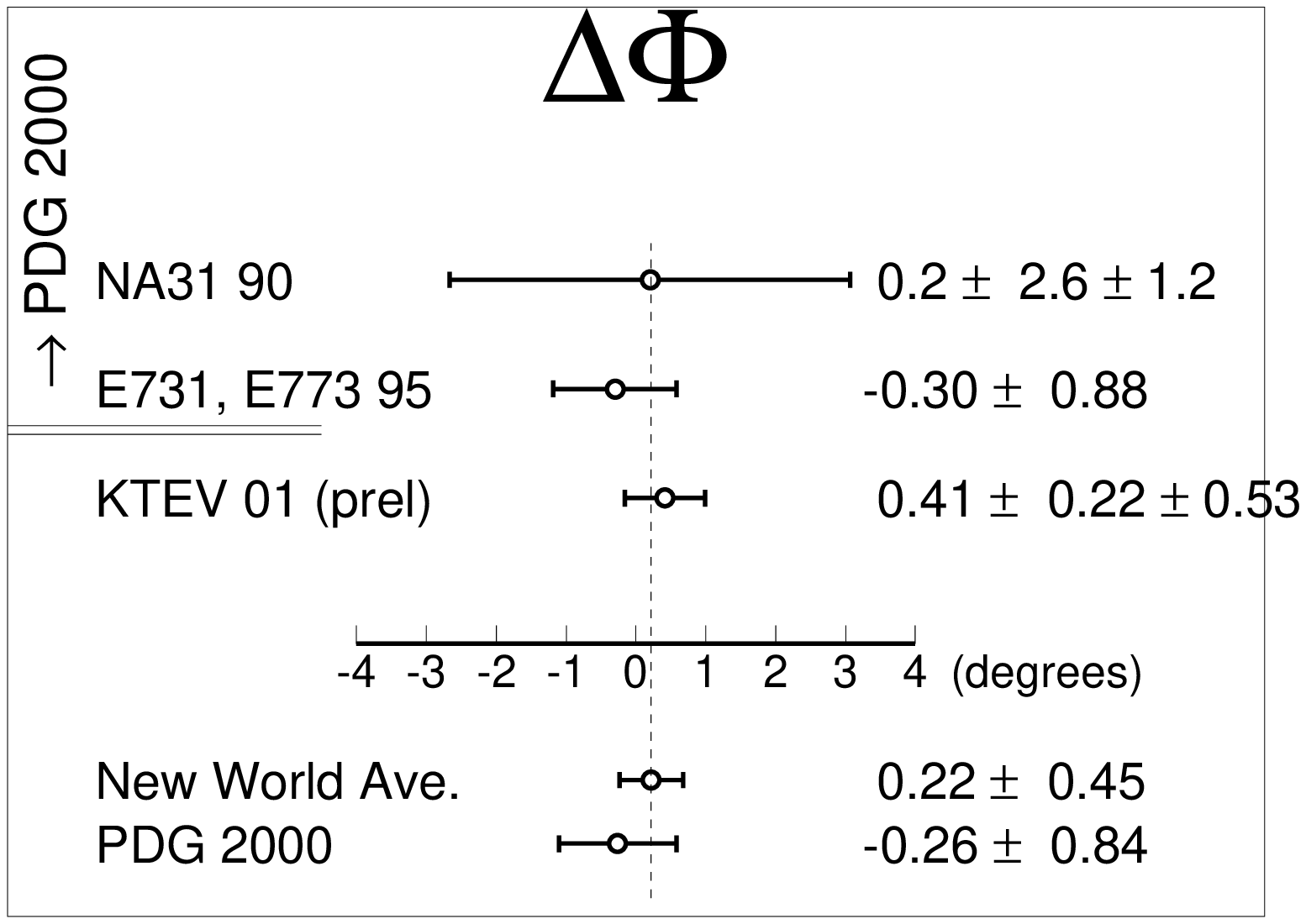}
\caption{History of $\delphi$ measurements.}
\label{fig:delphi}
\end{figure}

CPT-symmetry demands that 
\begin{equation}
 \delphi = \phipm - \phizz \simeq 0 ~.
\end{equation}
with the caveat that final state interactions can lead to 
$\delphi~\sim~0.05^{\circ}$.
In this fit, the uncertainty in $\phi_{\rho}$ cancels between the
charged and neutral phases.
The result from combining 96+97 data is 
\begin{equation}
      \delphi = [0.41 \pm 0.22_{stat} \pm 0.53_{syst} ]^{\circ}
\end{equation}
and is compared with previous results in Figure~\ref{fig:delphi}. 
Our value is consistent with CPT-symmetry.
The systematic uncertainty of $0.53^{\circ}$ is more than 
twice the statistical uncertainty, and is due mainly to the
neutral energy reconstruction.  Our $\delphi$ result is related to 
$\imepe$ by
\begin{eqnarray}
   \imepe & = & -\delphi/3 \nonumber \\
          & = & [-24 \pm 13_{stat} \pm 31_{syst} ]\eu \nonumber
\end{eqnarray}
Note that the uncertainty on $\imepe$ is $\times 12$ larger than
the uncertainty on $\epe$ (next section).

\begin{figure}[hb]
\epsfxsize200pt
\figurebox{160pt}{160pt}{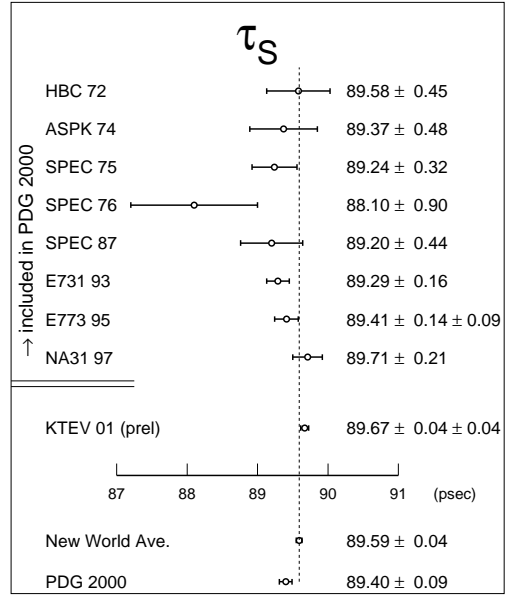}
\caption{History of $\tau_S$ measurements.}
\label{fig:taus}
\end{figure}

\begin{figure}[hb]
\epsfxsize200pt
\figurebox{160pt}{160pt}{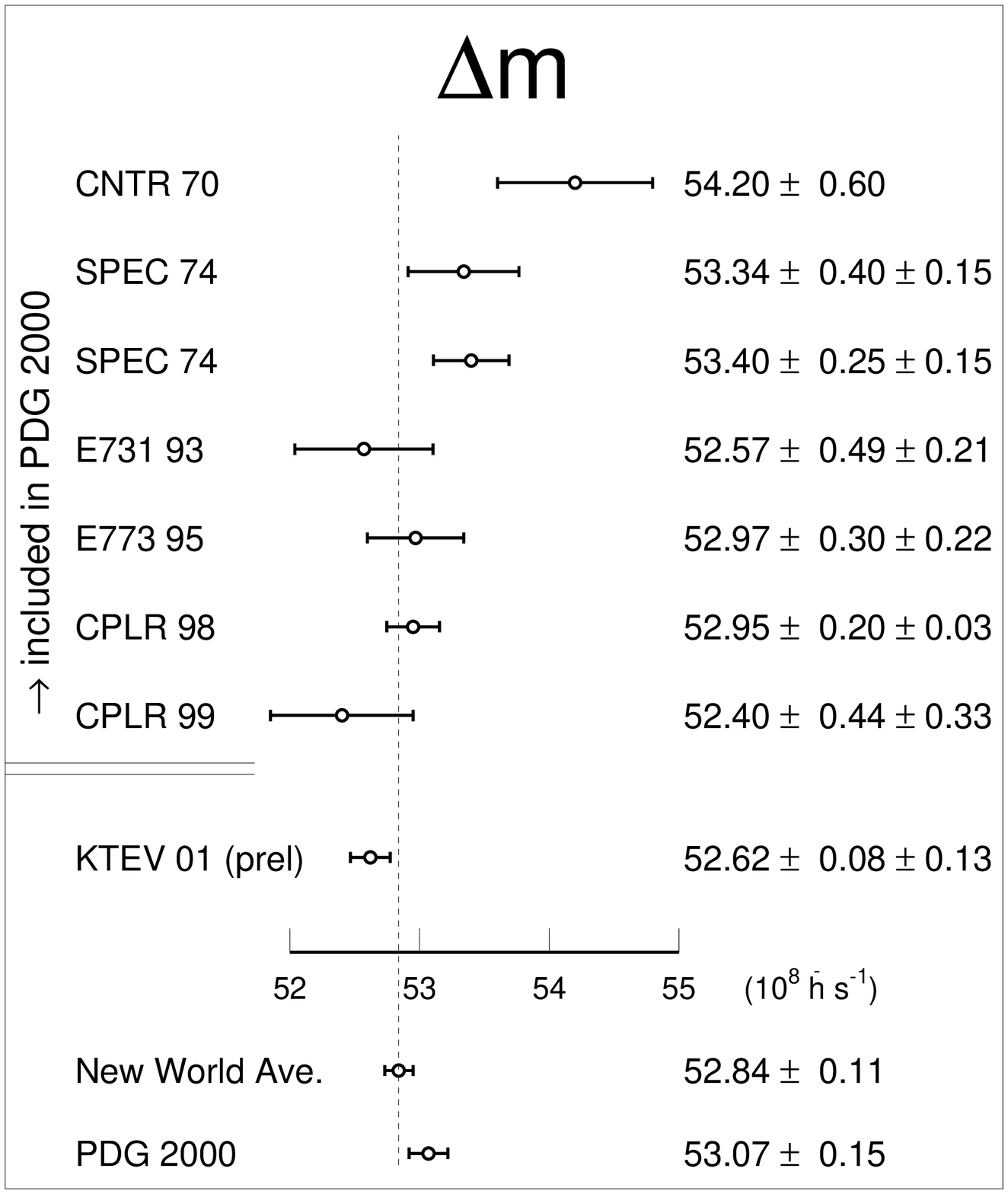}
\caption{History of $\delm$ measurements.}
\label{fig:delm}
\end{figure}

CPT symmetry also demands that $\phipm = \Phi_{SW}$.
From PDG2000, $\phipm - \Phi_{SW} = [-0.2\pm 0.5]^{\circ}$.
However, note that previous experiments had fixed 
$\tau_S$ and/or $\delm$ to PDG values in their fits;
our KTEV data show that these kaon parameters differ by more 
than $2\sigma$ from PDG values (Figures~\ref{fig:taus}-\ref{fig:delm}).
Since our measurements of $\tau_S$ and $\delm$ assume CPT-symmetry,
we cannot use our values to make a CPT test; on the other hand,
the PDG values may not be appropriate either. We therefore float
both $\delm$ and $\tau_S$ in our fit at the expense of increasing
both the statistical and systematic errors. 
Our preliminary result floating both $\tau_S$ and $\delm$ is
\begin{equation}
      \phipm - \Phi_{SW} = [+0.6 \pm 0.6_{stat} \pm 1.1_{syst}]^{\circ}
      \label{eq:phipm}
\end{equation}
An external measurement of $\tau_S$ that does NOT assume CPT would
help to reduce the KTEV errors as follows,
\begin{itemize}
   \item $\sigma_{stat}(\phipm-\Phi_{SW}) \to 0.3^{\circ}$
   \item $\sigma_{syst}(\phipm-\Phi_{SW}) \to \\ 0.7^{\circ}
                              \oplus 
           \left[ 7.3^{\circ} \times \sigma_{\tau_S}(psec)\right]$
\end{itemize}
If an external CPT-independent $\tau_S$ measurement has the same
error as KTEV (0.06~psec), then our error on 
$\phipm - \Phi_{SW}$ would be reduced from 
$1.3^{\circ}$ (Eq.~\ref{eq:phipm}) down to $0.9^{\circ}$.
There are good prospects for such an external measurement from the
CERN-NA48 collaboration.

\begin{figure}
\epsfxsize220pt
\figurebox{160pt}{160pt}{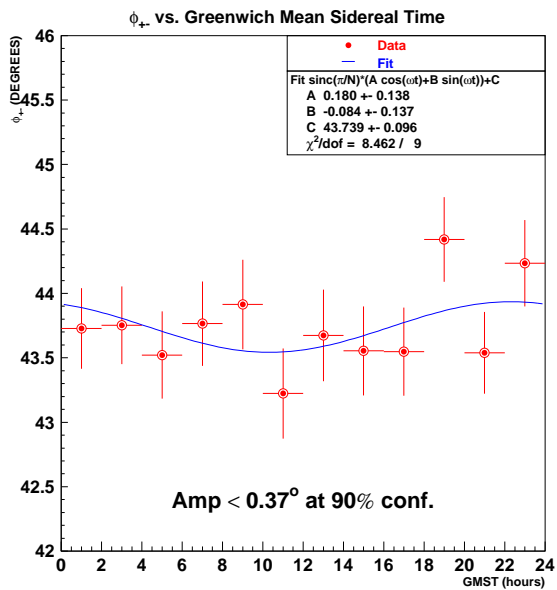}
\caption{$\phipm$ vs. sidereal time.}
\label{fig:diurnal}
\end{figure}

We have made another CPT test based on a suggestion 
to look for diurnal variations in $\phipm$\cite{alanK}.
The fit-values of $\phipm$ vs. sidereal time are shown in 
Figure~\ref{fig:diurnal}, which shows that $\phipm$
is constant over ``kaon beam direction'' to within $0.37^{\circ}$ 
at 90\% confidence.
A similar fit to $\tau_S$ limits diurnal variations
to be less than $0.0015\times \tau_S$

  \clearpage
  \section{$\epe$ Results}

\begin{figure}
\epsfxsize200pt
\figurebox{160pt}{160pt}{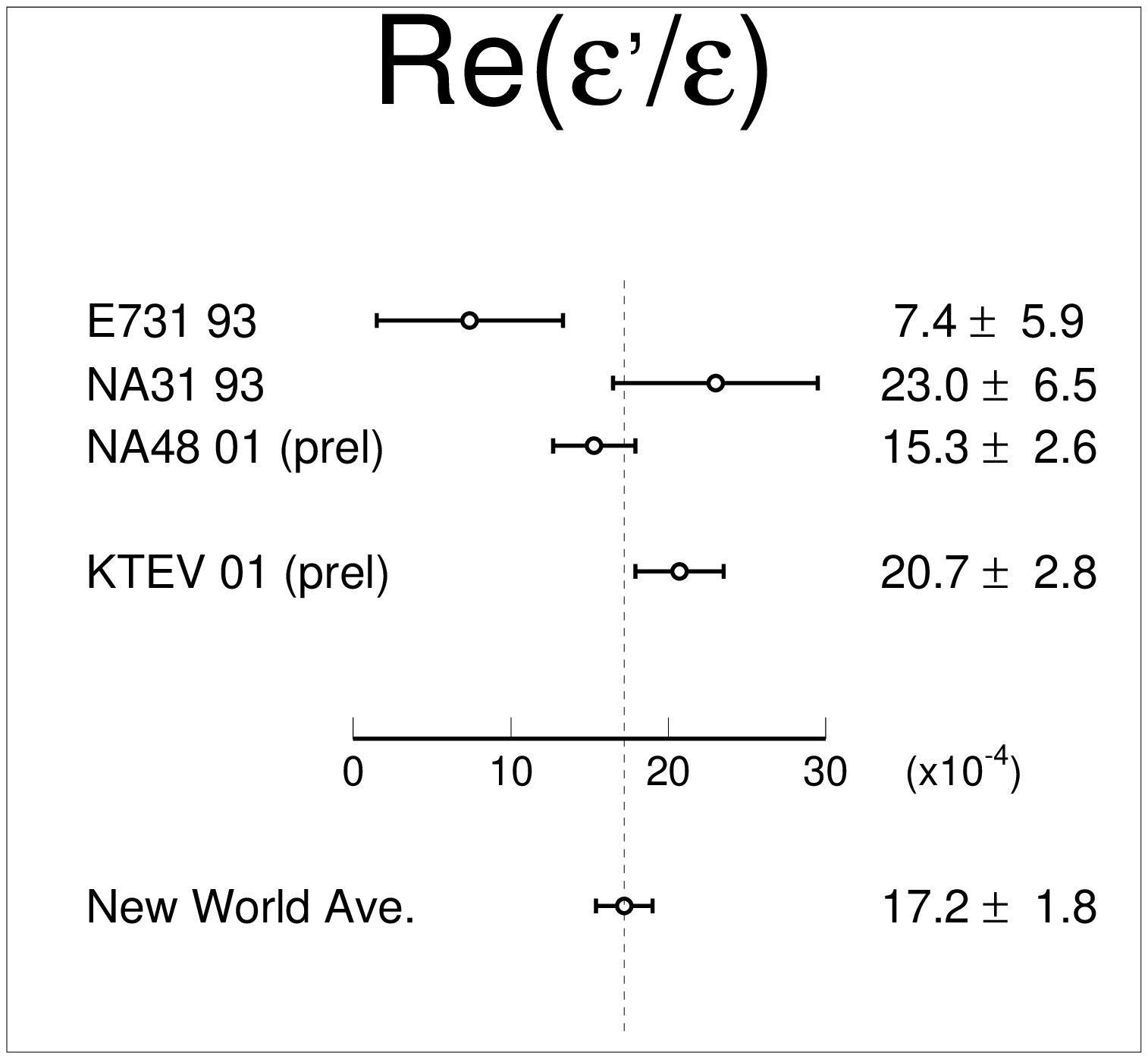}
\caption{$\epe$ results and new world average.}
\label{fig:epe}
\end{figure}
\begin{figure}
\epsfxsize230pt
\figurebox{160pt}{160pt}{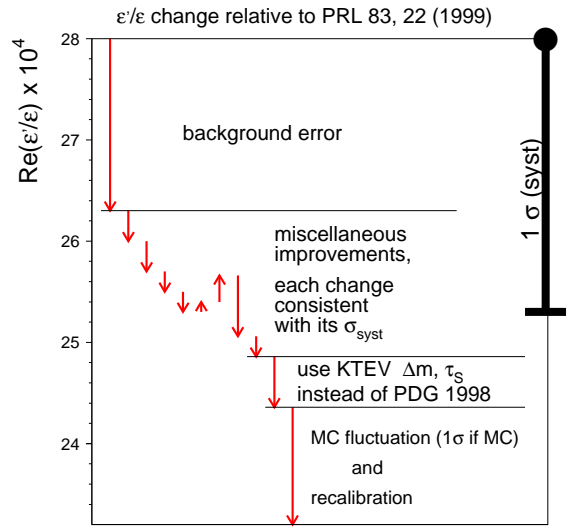}
\caption{
      Graphical illustration of changes to published result.
      The vertical axis shows $\epe$. The black bar at right
      shows the total $\sigma_{syst}$ 
      in the published result.
              }
\label{fig:dumarrow}
\end{figure}
\begin{figure}
\epsfxsize210pt
\figurebox{160pt}{160pt}{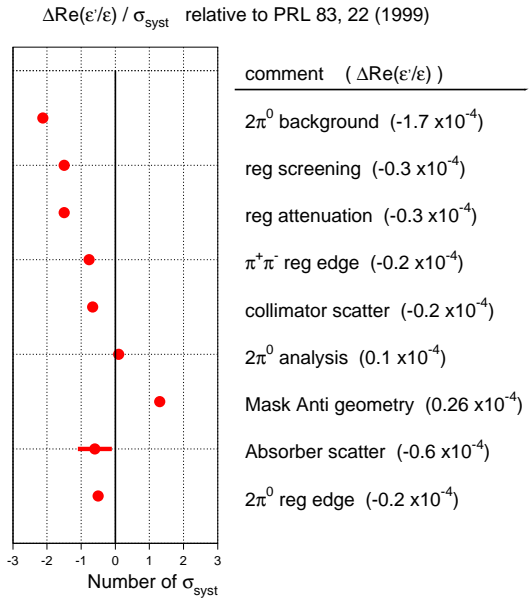}
\caption{Detailed list of $\epe$ improvements to the published result.
         The plot at left shows the number of systematic $\sigma$ 
         for each change. The numbers in () at the right show the
         corresponding change to $\epe$.  
        }
\label{fig:dumsyst}
\end{figure}

Our new and updated results on $\epe$ are shown in Table~\ref{tb:epe},
and a comparison with other results is shown in Figure~\ref{fig:epe}.
Using many improvements developed since the original result,
the $\epe$ update of the published sample\cite{epe99} has changed by 
$-4.8\eu$, and the changes are illustrated in 
Figures~\ref{fig:dumarrow}-\ref{fig:dumsyst}.
The changes are due to analysis improvements, better measurements
of $\tau_S$ and $\delm$, and to MC statistical fluctuations.
Except for a mistake in the {\reg}-scatter background
(top entry in Fig.~\ref{fig:dumsyst}),
the changes are consistent with the systematic errors assigned
in the published result. 
All of the changes to $\epe$ are {\it uncorrelated}.
The updated result has a slightly larger systematic error,
$3.2\eu$ compared with published value of $2.7\eu$;
this is because the data/MC decay vertex comparisons, which determine
the acceptance error, are slightly worse using the improved techniques, 
even though the larger independent [1997] data set shows much 
better agreement in the same distributions.

\begin{table}
\caption{ Summary of KTEV $\epe$ results. The 1997 and ``PRL'' samples
          are statistically independent.  The uncertainties refelct
          statistical, systematic and MC-statistics.
        } 
\label{tb:epe}
\begin{tabular}[t]{|c|c|}
\hline
  data      & {\it PRELIMINARY}        \\
  sample    & $\epe (\eu)$              \\
 \hline
  1997     & $19.8 \pm 1.7_{stat} \pm 2.3_{syst} \pm 0.6_{MC}$ \\
           &                                                        \\
  ``PRL''  & $23.2 \pm 3.0_{stat} \pm 3.2_{syst} \pm 0.7_{MC}$ \\
  (update) &                                                        \\
\hline 
  96+97 &  $20.7 \pm 1.5_{stat} \pm 2.4_{syst} \pm 0.5_{MC}$ \\
\hline
\end{tabular} 
\end{table}

      \subsection{$\epe$ Systematic Errors}

The dominant systematic errors are shown in Table~\ref{tb:epe syst}.
The neutral energy uncertainty is determined  mainly from $\pzpz$ pairs
produced from hadronic interactions in the vacuum window located 
159 meters from the primary target. The vacuum window location is known
to within 1~mm based on charged track vertex analyses; 
the $\pzpz$ vertex is off by $2.5\pm 0.4$~cm, 
leading to most of the error in the first row
of Table~\ref{tb:epe syst}.

The neutral {\bkg} from {\reg} scatters is based on measuring
the $\pt2$ distribution from $\kchrg$ decays in the {\reg} beam, 
and then using this
measurement to simulate $\kneut$ decays that scatter in the {\reg}.
The $\sim 1$\% scatter {\bkg} in neutral mode must be known to 
about 5\% of itself to get the $\epe$ systematic down to $1\eu$.
The {\bkg} uncertainty is due to the charged mode reconstruction and
acceptance ($0.8\eu$), to the possibility that the beam-hole veto
used only in neutral could alter the $\pt2$ shape ($0.3\eu$)
and to slight imperfections in the parametrization of the 
$\pt2$ using scattered $\kchrg$ decays  ($0.4\eu$).

The charged acceptance uncertainty in row 3 of Table~\ref{tb:epe syst}
is due mainly to a large data/MC $z$-slope in the first 20\% of the
1997 data set. The $\epe$ charged acceptance error is $2.2\eu$ in the 
first 20\% of the data and $0.5\eu$ for the remaining 80\%.

During data collection, a Level 3 software filter was used to 
remove most of the  $\ke3$ and $K_L\to\pi\mu\nu$ events.
One percent of the charged triggers were saved without any online filter 
to check for a bias in $\kchrg$ decays. 
We find a small bias with $2.2\sigma$ significance,
which leads to a systematic error of $0.5\eu$ on $\epe$.
More detailed studies on the higher statistics sample of un-filtered
$\ke3$ decays shows no Level 3 bias in the decay vertex distribution, 
nor any difference in loss between the two beams.

The neutral apertures consist of a collar-anti veto [CA] surrounding the
CsI beam-holes, a mask-anti veto just upstream of the {\reg},
and the CsI size. The aperture-systematic is dominated by the
$100~\mu m$ uncertainty in the size of the CA. Note that the CA size
was determined in-situ using electrons from $\ke3$ decays.

\begin{table}
\caption{Dominant $\epe$ systematic uncertainties ($\eu$)}
\label{tb:epe syst}
\begin{tabular}[t]{| l c |}
\hline
    Neutral Energy      &  1.5  \\  
    reconstruction      &       \\
 \hline
    Neutral background from        &  1.0   \\
    reg-scatters (B/S$\sim$ 1\%)   &        \\
 \hline
    Charged Acceptance  &  0.9   \\ 
    (data/MC $z$-slope) &        \\       
 \hline
    Charged Level 3 online        &  0.5 \\ 
    filter ($2.2\sigma$ effect)   &      \\
 \hline
    Neutral apertures             &  0.5 \\ 
 \hline
\end{tabular}
\end{table}

      \subsection{$\epe$ From Re-weighting Method}

\begin{figure}
\epsfxsize180pt
\figurebox{160pt}{160pt}{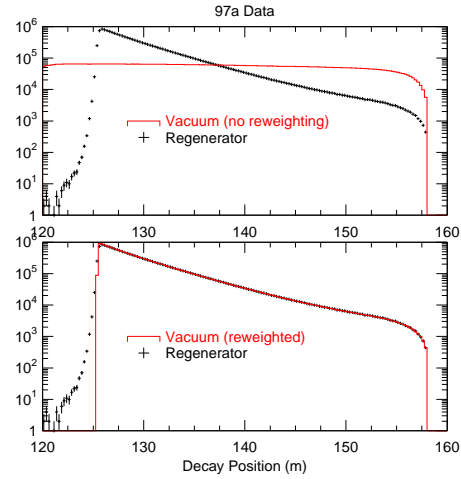}
\caption{
     Decay distributions for vacuum $(K_L$) and {\reg} $(K_S)$
     before (top) and after (bottom) re-weighting.
        }
\label{fig:zrewgt}
\end{figure}

\begin{figure}
\epsfxsize180pt
\figurebox{160pt}{160pt}{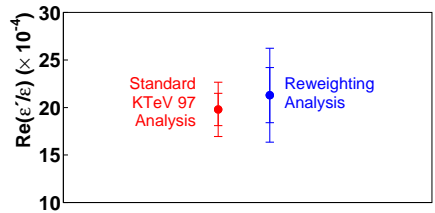}
\caption{
     {\it PRELIMINARY} comparison of Standard $\epe$ result 
     using MC correction
     with re-weight analysis that does NOT use MC correction.
        }
\label{fig:eperewgt}
\end{figure}

As an additional cross-check, $\epe$ was measured using
a re-weighting technique that is similar to the NA48 method, 
except that no MC correction is needed due to the identical 
phase space of the two kaon beams in the KTEV experiment. 
The method is illustrated by the decay 
distributions shown before and after re-weighting 
in Figure~\ref{fig:zrewgt}. After re-weighting the acceptances
are the same in the two beams.
A comparison of the nominal and re-weighting result using 1997 data
sample (PRL sample is not included) is shown in 
Figure~\ref{fig:eperewgt}. The difference between the two analyses
is
\begin{equation}
   \Delta\epe = [1.5\pm 2.1_{stat} \pm 3_{syst}] \eu  
\end{equation}
where the systematic uncertainty is dominated by
cut-variations, particularly on the minimum cluster
energy in the $\kneut$ analysis.

  \section{$\ke3$ Charge Asymmetry}

\begin{figure}
\epsfxsize180pt
\figurebox{160pt}{160pt}{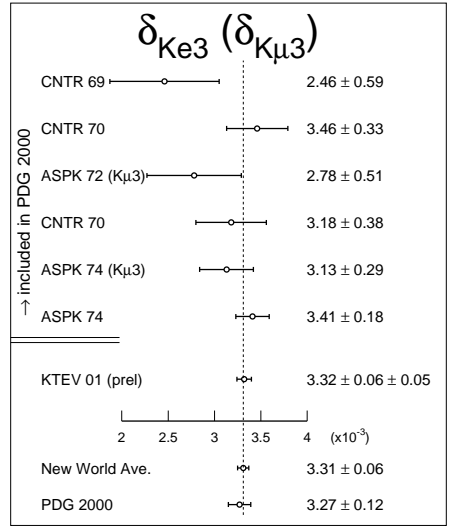}
\caption{
     History of $\dke3$ and $\delta_{K\mu 3}$ results.
        }
\label{fig:dke3}
\end{figure}

The $\ke3$ Charge Asymmetry is defined to be
\begin{eqnarray}
   \dke3 &  \equiv &
         {  {N(\pi^- e^+ \nu) - N(\pi^+ e^- \nu) }
                     \over 
            {N(\pi^- e^+ \nu) + N(\pi^+ e^- \nu) }
         }                   \label{eq:dke3} \\
         &   &      \nonumber \\
         & = &  2Re(\epsilon- \Delta - Y -  X_{-})
                    \nonumber 
\end{eqnarray}
where
\begin{eqnarray}
   \epsilon  & = & {\CPV}~in~mixing.   \nonumber \\
   \Delta    & = & {\CPTV}~in~mixing~(\Delta S = \Delta Q).   
                    \nonumber \\
    Y        & = &  {\CPTV}~in~decay~amplitude~(\Delta S = \Delta Q).
                    \nonumber \\
    X_{-}    & = &  {\CPTV}~in~decay~amplitude~(\Delta S \ne \Delta Q).
                \nonumber
\end{eqnarray}
The previous best measurement was made at CERN in 1974, and was
based on 34 million events. KTEV has a new measurement based on
298 million decays in the $K_L$ beam ({\reg} decays are not used).  
Each $N$-factor in Eq.~\ref{eq:dke3} is replaced with the four-fold 
geometric mean of the two beams (left,right) and the two magnet
polarities in order to cancel acceptances. There is no MC acceptance
correction, but there are corrections for particle/antiparticle 
differences. These differences are measured in-situ using 
complementary samples such as $K_L\to\pipi\pi^0$ to study 
the $\pi^+$ and $\pi^-$ detection efficiencies.
Our new result is 
  $$ \dke3 = [ 3.32 \pm 0.06_{stat} \pm 0.05_{syst}] \times 10^{-3} $$
and is shown in Figure~\ref{fig:dke3} with previous measurements.
The world average uncertainty is reduced by a factor of 2
with the addition of the KTEV measurement.

CPT limits can be obtained from the relation,
\begin{equation}
   \Re( {2\over 3}\eta_{+-} + {1\over 3}\eta_{00} ) - \delta_L
      = \Re(Y + X_{-} + a)
\end{equation}
where  $\delta_L$ is the average of $\dke3$ and $\delta_{K\mu 3}$
asymmetries, $Y$ and $X_{-}$ are defined above and $\Re(a)$ 
parametrizes $\CPTV$  in  $K (\kbar)\to 2\pi$ decays.
The result is that the sum of $\CPTV$ parameters is
\begin{equation}
     \Re(Y + X_{-} + a) = (-3 \pm 35) \times 10^{-6}
\end{equation}
which is less than $55\times 10^{-6}$ with 90\% confidence.

  \clearpage
  \section{Rare Kaon Decays}

\begin{figure*}
\epsfxsize400pt
\figurebox{160pt}{160pt}{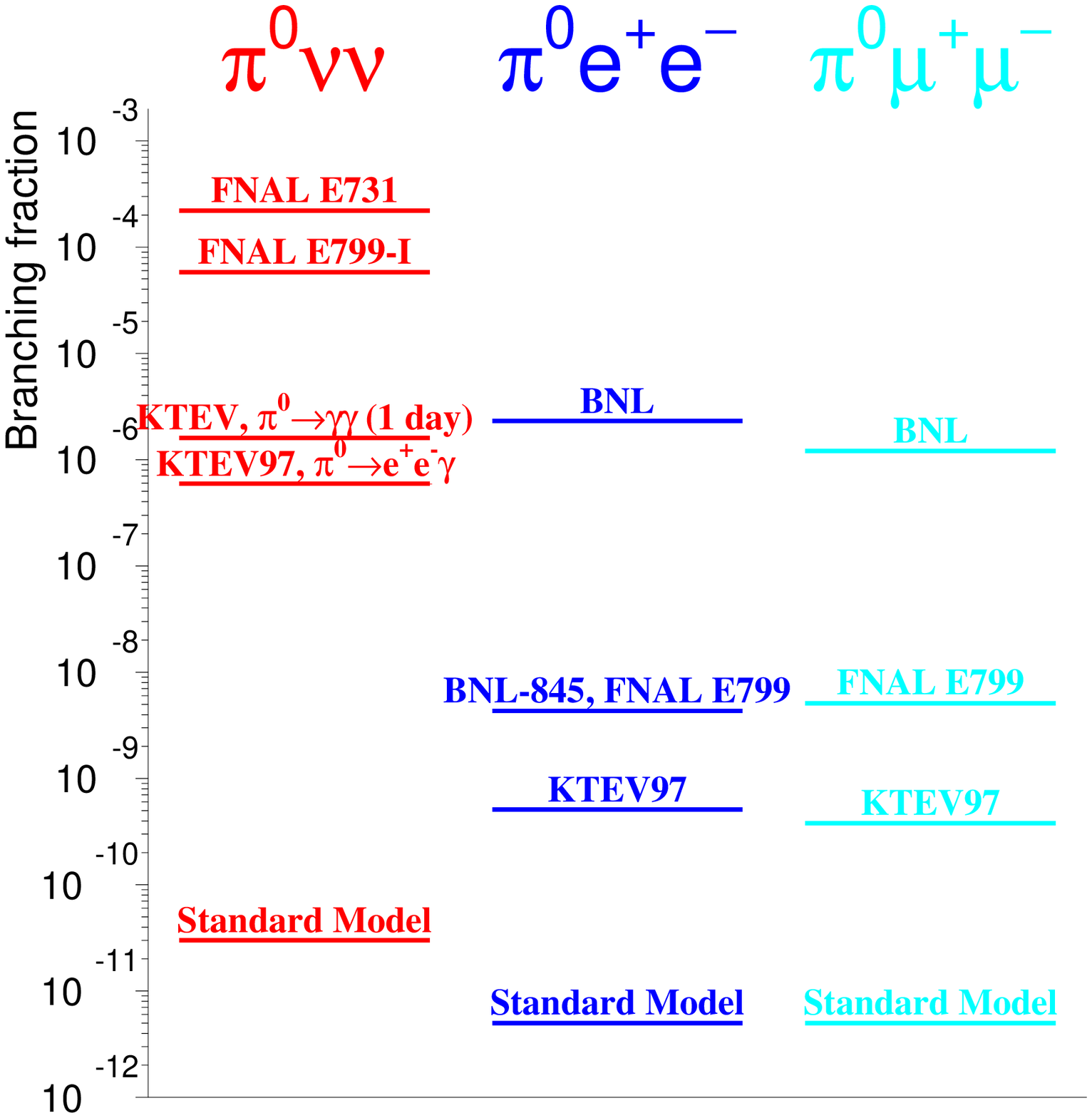}
\caption{
     History of $K_L\to\pi^0\llbar$ results.
        }
\label{fig:rarecpvdk}
\end{figure*}

Table~\ref{tb:raredk} lists more than a dozen rare decay results 
that have been improved by the KTEV collaboration.
Typical improvements are based on order-of-magnitude increases
in statistics or sensitivity, and includes three first-observations.
In the $\CPV$ column we have
1800 $K_L\to\pp\ee$ events which shows a 13\% $\CPV$ asymmetry
in the angle between the $\pp$ and $\ee$ planes\cite{ppee}.
The prospects for observing direct $\CPV$ in $K_L\to\pi^0\llbar$ modes
($\pi^0\nu\bar{\nu},~\pi^0\ee~{\rm and}~\pi^0\mm$)
requires measuring BRs down to $\sim 10^{-11}$ (Fig.~\ref{fig:rarecpvdk}).
In the case of $K_L\to\pi^0\nn$, KTEV has improved the upper limit
by $\sim 10^2$, but is still a factor of $10^4$ shy of the
standard model.  KTEV has improved the $K_L\to\pi^0\ll$ modes
by a factor of 10, and remains a factor of $10^2$ above the
standard model predictions.

\begin{table*}
\caption{
     List of rare kaon decay modes by physics topic for which
     KTEV has made improved BR measurements. The numbers in 
     parentheses indicate the increase in statistics or sensitivity.
     ``\firstobs'' indicates first observation.
        }
\label{tb:raredk}
\begin{tabular}{| l | l | l | l | } \hline
           &             &             &  lepton      \\
           &  electromag. & hadronic   &  flavor      \\
 \CPV      &  structure   &  structure &  violation   \\
 \hline\hline
 $\pp\ee$   \firstobs   & $\mm\ee$     \kgain{38}    &   
 $\pi^0\gg$ \kgain{14}  & $\pi^0\mu e$ \kgain{14}  \\
 $\pi^0\ee$   \kgain{8}      &  $\ee\ee$ \kgain{7}   &
 $\pi^0\ee\gamma$ \firstobs  &               \\
 $\pi^0\mm$ \kgain{13}   &  $\mm\gamma$ \kgain{39}  & 
 $\pzpz\ee$              &               \\
 $\pi^0\nu\bar{\nu}$ \kgain{100} & $\ee\gg$ \kgain{15}  &
                                 &               \\
 $\pp\gamma$ \kgain{2}   & $\mm\gg$ \firstobs    &
                         &               \\
\hline
\end{tabular}
\end{table*}

\end{document}